\documentclass[12pt,preprint]{aastex}
\begin{document}
\newcommand{\msun}{\mbox{M$_{\odot}$}}
\newcommand{\rsun}{\mbox{R$_{\odot}$}}
\title{Detection of a Variable Infrared Excess Around SDSS 121209.31+013627.7\altaffilmark{1}}

\author{J.H. Debes, M. L\'opez-Morales\altaffilmark{2}, A.Z. Bonanos\altaffilmark{3}, A.J. Weinberger}
\affil{Carnegie Institution of Washington, Department of Terrestrial Magnetism,\\
5241 Broad Branch Road, Washington, DC 20015}
\affil{\tt e-mail:~debes,mercedes,bonanos,weinberger@dtm.ciw.edu}

\altaffiltext{1}{Based on observations obtained with the 2.5 m DuPont
telescope and the 6.5 m Magellan Telescopes at Las Campanas Observatory, which is operated by the Carnegie
Institution of Washington.}
\altaffiltext{2}{Carnegie Fellow.}
\altaffiltext{3}{Vera Rubin Fellow.}

\begin{abstract}
We present near-infrared $JHK_s$ photometry and light curves of
the candidate magnetic white dwarf+brown dwarf binary SDSS
J121209.31+013627.7 and report on the detection of near-infrared excess
and variability in the $K_s-$band. The observed near-infrared excess can
be explained by the presence of an L7 brown dwarf and an extra emission
source.  The $J$ and $H$ light curves appear flat, which rules out eclipses
deeper than 0.2 mag and the presence of an accretion hot spot on the
white dwarf.  From the variable $K_s$ lightcurve, we get a refined period for
the system of 88$\pm$1 minutes.  We show that the observed variability in $K_s-$band can be explained by cyclotron
emission, which can be modeled by a small spot on the surface of the
white dwarf.
SDSS 1212 exhibits similarities to the ultra-short period polar EF
Eridani, however the lack of evidence for Roche-lobe overflow accretion
suggests it may be a pre-polar.

\end{abstract}


\section{Introduction}

SDSS J121209.31+013627.7 (hereafter SDSS 1212) was first reported by
\citet{schmidt03} as a magnetic white dwarf with an equivalent dipolar
magnetic field of $B_{d}$ = 13~MG. Recently, \citet{schmidt05} published
a new study suggesting that SDSS 1212 is a white dwarf/probable brown
dwarf binary. Their follow-up spectroscopy revealed the presence of a
narrow H$\alpha$ emission line with a semi-amplitude radial velocity
variation of $320\pm20$~km~$s^{-1}$ and an orbital period of
$93.6\pm14.4$ minutes, indicating the presence of a nearby irradiated
companion. The H$\alpha$ emission disappeared when the companion faced the
white dwarf, suggesting a high inclination orbit.  From the Zeeman splitting of
hydrogen absorption lines in the photosphere, \citet{schmidt05} measured
a mean magnetic surface field of 7~MG. Their $J-$band photometry yielded
$17.91\pm0.05$ mag, placing an upper limit on the mass of the companion
and thus suggesting the brown dwarf scenario.

In this $Letter$, we present near-infrared $JHK_s$ observations of SDSS
1212, aimed at detecting the brown dwarf companion and determining the
presence or absence of eclipses of the white dwarf. 

\section{Observations and Data Analysis}
\label{sec:obs}
We observed SDSS 1212 in $JHK_s$ over five nights, 2006 February 15 and 17
UT, with the WIRC camera on the 2.5 m DuPont
telescope and from 2006 June 4 to June 7 UT with the PANIC camera on the 6.5 m Baade Telescope at Las Campanas Observatory, Chile. The near-infrared filter set at Las Campanas (LCO system) is
 described
by \citet{persson98}.

With WIRC, we observed SDSS~1212 with $\sim$1 minute sampling in all 
three filters, observing in $J$ for 84 minutes and $H$ for 15 minutes.
The second night we observed in $J$ and $K_s$ for 80 minutes and in $H$ for
15 minutes. 
Our PANIC observations sampled SDSS~1212 every 30~s
 for all three filters, with observations
spanning $\sim$120 minutes for each filter.  
Typical seeing over the five nights ranged from 
0$\farcs$6 to 1$\farcs$3. 
We used a five point dither pattern for each data set.

The images were processed using standard IRAF\footnote{IRAF is
distributed by the National Optical Astronomy Observatories, which are
operated by the Association of Universities for Research in Astronomy,
Inc., under cooperative agreement with the NSF.} routines or equivalent IDL programs. Each frame was flat-fielded, background subtracted, and bad-pixel corrected.
 We then performed simultaneous
differential aperture photometry on each image of SDSS 1212 and
comparison stars in the field. All of the
comparisons behaved similarly, showing no individual photometric
variation trends over the time span of the observations. 
We derived differential magnitude light curves for
each comparison using aperture sizes that were approximately equal to
the width of the seeing and averaged them to obtain the final light
curves. 

We used six comparison stars in the images. 
The 2MASS\footnote{The Two Micron All Sky
Survey \citep[2MASS][]{cutri03,skrutskie06} is a joint project of the
University of Massachusetts and the Infrared Processing and Analysis
Center/California Institute of Technology, funded by the National
Aeronautics and Space Administration and the National Science
Foundation.} apparent magnitudes of those
stars are summarized in Table \ref{tab:photometry}.  Our dithering
pattern precluded all six stars from being present in every frame.
The PANIC data had three of the comparison 
stars, due to its smaller field of view.  
For each night we took two bright comparison
stars in the field to
calculate the absolute photometry of SDSS 1212, based on their 2MASS
photometry. The final $JHK_s$ photometry is presented in
Table~\ref{tab:stub}. Transforming the comparison star 2MASS photometry
into the LCO system gives values that are the same within the
photometric errors \citep{carpenter01}.

\section{Results}
\label{sec:results}
\subsection{$JHK_s$ Light Curves}
\label{subsec:lightcurve}

The light curves of SDSS 1212 were folded using the ephemeris equation
\begin{equation}
\label{eq:eph}
\rm T(HJD) = 2453783.8375 + 0.061 \times E,
\end{equation}

\noindent where we have adopted the orbital period of 0.061$\pm$0.001 days 
derived by finding a period that minimized the phased point-to-point
variations in our $K_s$
lightcurves \citep{lafler65}.  We have chosen the mid-point of the magnitude dip
observed in $K_s-$band (see below) as phase 0. The phased light curves 
from all nights are shown in Figure \ref{fig:lightcurve}.

The $J$ and $H$ light curves appear featureless up to the one standard
deviation precision level of our photometry ($\sim$ 0.07 mags for $J$ 
and $H$), which
precludes eclipses deeper than 0.2 mag. However, there is a marginally detected dip in the $H$ curve at a phase of 0.5 which appears in both
nights of data.  The emission of SDSS 1212 in
$J-$band is dominated by the white dwarf, therefore, if eclipses were
present, one would clearly observe the primary eclipse (occultation of
the white dwarf by the companion), 
but the eclipse of the secondary
would still be undetectable. A white dwarf with a typical radius $\sim
0.012\; \rm R_{\sun}$ in an edge-on ($i$ = 90 deg) configuration with a
$\sim 0.11 \;\rm R_{\sun}$ brown dwarf would produce a $\sim$5 min
eclipse of the white dwarf. Using the most recent version of the
Wilson-Devinney program \citep[WD2003,][]{wilson71,wilson79}, we
generated model light curves for a range of inclinations and radii of
the secondary, keeping the white dwarf radius and the orbital separation
fixed, and compared the results to the light curve in $J-$band to
conclude that the inclination of the system is constrained to $i$
$\lesssim 78.5$ deg.

The $K_s-$band light curve is
marked by a flat dip which lasts for $\sim$ 41 minutes. That
dip is followed by a brightening of the system by $\sim$ 0.7 mag, 
with a second dip at a phase of 0.5.

The light curve variations of SDSS 1212 in $K_s-$band are
reminiscent of those detected in the ultrashort period
($P_{orb}=81$ minutes) polar EF Eri by \citet{harrison03}. Those
variations were initially attributed to a cool irradiated companion with
the side facing the white dwarf at a higher temperature. However, that
interpretation was refuted in \citet{harrison04} after they failed to
detect phase-dependent spectral line changes that an irradiated
companion would produce in the infrared spectrum of the system. They
concluded that cyclotron emission is responsible for the photometric
variability observed in EF Eri. In addition, the $J-$band light
curve of EF Eri presents a $\sim$ 0.5 mag sinusoidal variation that
has been attributed to a bright spot on the white dwarf.  As of
10 March 2006, EF Eri returned to its high state 
\citep{howell06}.

However, the SDSS 1212 light curves differ from those of EF Eri. There is 
no sign of sinusoidal variation in $J-$band, which
indicates the absence of an accretion spot on the white dwarf. The
$K_s-$band light curves show a sharp drop in magnitude followed by the
flat dip, unlike the sinusoidal changes observed in EF Eri. This
suggests that, whatever the source of the extra emission in $K_s-$band
is, it originates in a more localized region that is
being obscured during the dimmer phases. 

\subsection{Infrared Excess}
\label{subsec:IRex}

Given that the $J-$band light curve is flat, 
we derived an apparent magnitude of $J=17.90\pm0.06$ by
averaging over all phases. Similarly, there appears to be no significant
variability in the $H-$band, and we measure $H=17.56\pm0.05$~mag. In
$K_s-$band we derived two values for the total flux. The first value,
$K_s=16.53\pm0.08$~mag, corresponds to the average magnitude of the
system between phases $\phi = 0.2-0.4$ ($Stage$ $1$), where we are
detecting emission from the white dwarf, its companion, and any
additional radiation present in the system. The second value,
$K_s=17.23\pm0.09$ mag, was computed as the average magnitude between
phases $\phi = 0.0-0.2$ and $0.85-1.0$ ($Stage$ $2$) which
 corresponds to the white dwarf and the companion being visible.

\begin{figure}
\plotone{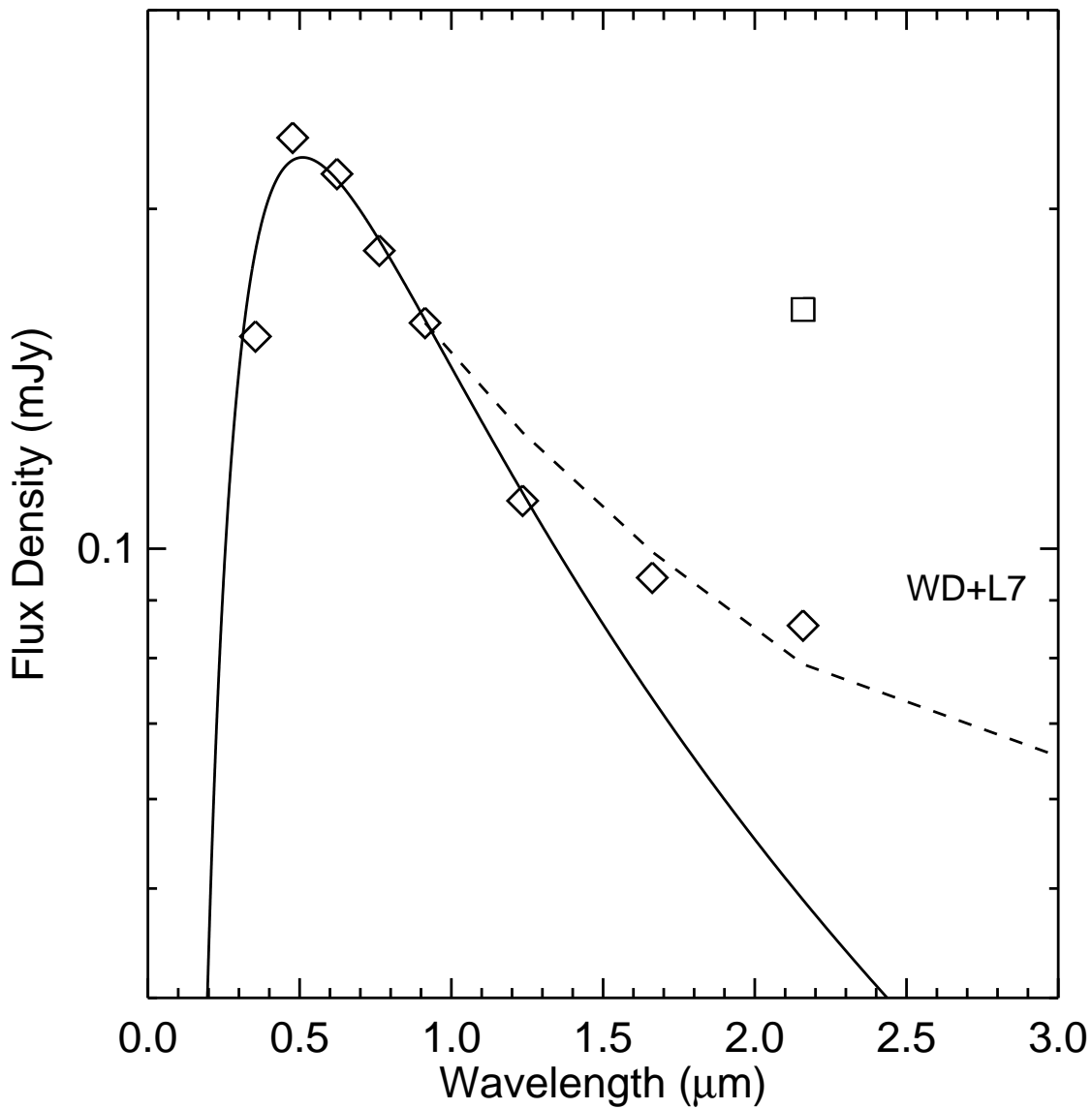}
\caption{\label{fig:sed} Spectral energy distribution of SDSS 1212. The
solid line represents the SED of a white dwarf at 10000K.  The dashed
line represents the white dwarf plus an 
L7 brown dwarf at 145~pc.  The points correspond to
the Sloan $ugriz$ photometry and the $JHK_s$ photometry presented in
Section \ref{sec:results}. Note that in the $K_s-$band there are two
values corresponding to $Stage 2$ (diamond) and $Stage 1$ (open
square).  Errors in $JHK_s$ are comparable to the symbol size.}
\end{figure}

The resultant spectral energy distribution (SED) of SDSS 1212 is shown
in Figure \ref{fig:sed}. The diamonds correspond to the average flux of
the system in the Sloan Digital Sky Survey (SDSS) $ugriz$ bands
\citep{fukugita96} and the $J$, $H$, and $K_s$ photometry in $Stage$
$2$, when SDSS 1212 is dimmer. The open square represents the average
$K_s$ flux of the system during the brighter phase of $Stage$ $1$.  

The flux of SDSS 1212 from the optical to $J$ is reproduced by a
blackbody model of a single white dwarf with an effective temperature of
about 10,000 K (see solid line in Figure \ref{fig:sed}), in agreement with the
conclusion reached by \citet{schmidt05}. Our $H$ and $K_s-$band data, on
the other hand, clearly show infrared excess in each one of the stages
considered above. Therefore, we conclude that there is a substellar
companion, most likely of middle to late L spectral type and with
T$_{eff}<1700$~K. The amount of infrared excess can be determined in
each case by subtracting the predicted white dwarf flux at each
wavelength band.  The main uncertainty in this procedure is the poorly
known temperature of the white dwarf.  To estimate the flux of the white
dwarf we use a $T_{eff}=10000\pm1000$~K \citep{schmidt05}.  We estimate
the formal error introduced into the measurement of the excess by taking
the differences in predicted flux at $JHK_s$ and scaling them to the
measured flux of the system at $J$.  In $J$, there is no significant
excess, so we take a 3$\sigma$ upper limit, where we calculate
$\sigma=\sqrt{\sigma_{J}^2+\sigma_{WD}^2}$.  The resultant apparent
magnitudes of the near-IR excess are J$>19.6$, $H$=19.2$\pm$0.2, and
$K_s$=16.9$\pm$0.1, for $Stage$ $1$, and 18.2$\pm$0.3 for $Stage$
$2$. 

In order to determine whether the excess in $Stage$ $2$ is consistent
with flux solely coming from the white dwarf's companion, we compare the
apparent magnitudes of the excess to the mean absolute magnitudes of
field brown dwarfs with measured parallaxes \citep{vrba04}. Assuming a
distance to SDSS 1212 of 145~pc results in an excess with absolute
magnitudes in $J, H$, and $K_s$ of $>$13.8, 13.4$\pm$0.2, and
12.4$\pm$0.3 respectively.  These absolute magnitudes match with a
spectral type of L7.  SDSS 1212's distance is uncertain by 14\% and is
based on theoretical modeling of the white dwarf's atmosphere.  This
uncertainty translates an error of $\pm$1 spectral type.  
Figure \ref{fig:sed} shows a
comparison of the observed SDSS~1212 photometry (diamonds) and the combination of the
expected WD fluxes and those from an L7 brown dwarf (dashed line).  
By definition, the model
doesn't fit the observed photometry in the $J$-band since we are assuming 
a 3$-\sigma$ upper limit.

\section{Discussion}
\label{sec:discussion}

There are two possible explanations for the SDSS~1212 lightcurve: (1) the
excess emission is due to a hotter irradiated side of the companion, or
(2) the excess is produced by a separate mechanism, such as cyclotron
emission.

In the case of irradiation, the SDSS 1212 system would appear brighter
during the phases when the hot side of the companion is facing towards
us, with its brightness dropping once per orbit as the hot side of the
companion faces away from us. The difference in flux between $Stage$ $1$
and $Stage$ $2$ is approximately a factor of 1.5. Assuming flux
from two pure blackbodies, one would expect the temperature ratio
between the hot and the cool sides of the companion to account for this
difference to be $T_{hot}/T_{cool} = 1.1$.

\begin{figure}
\plotone{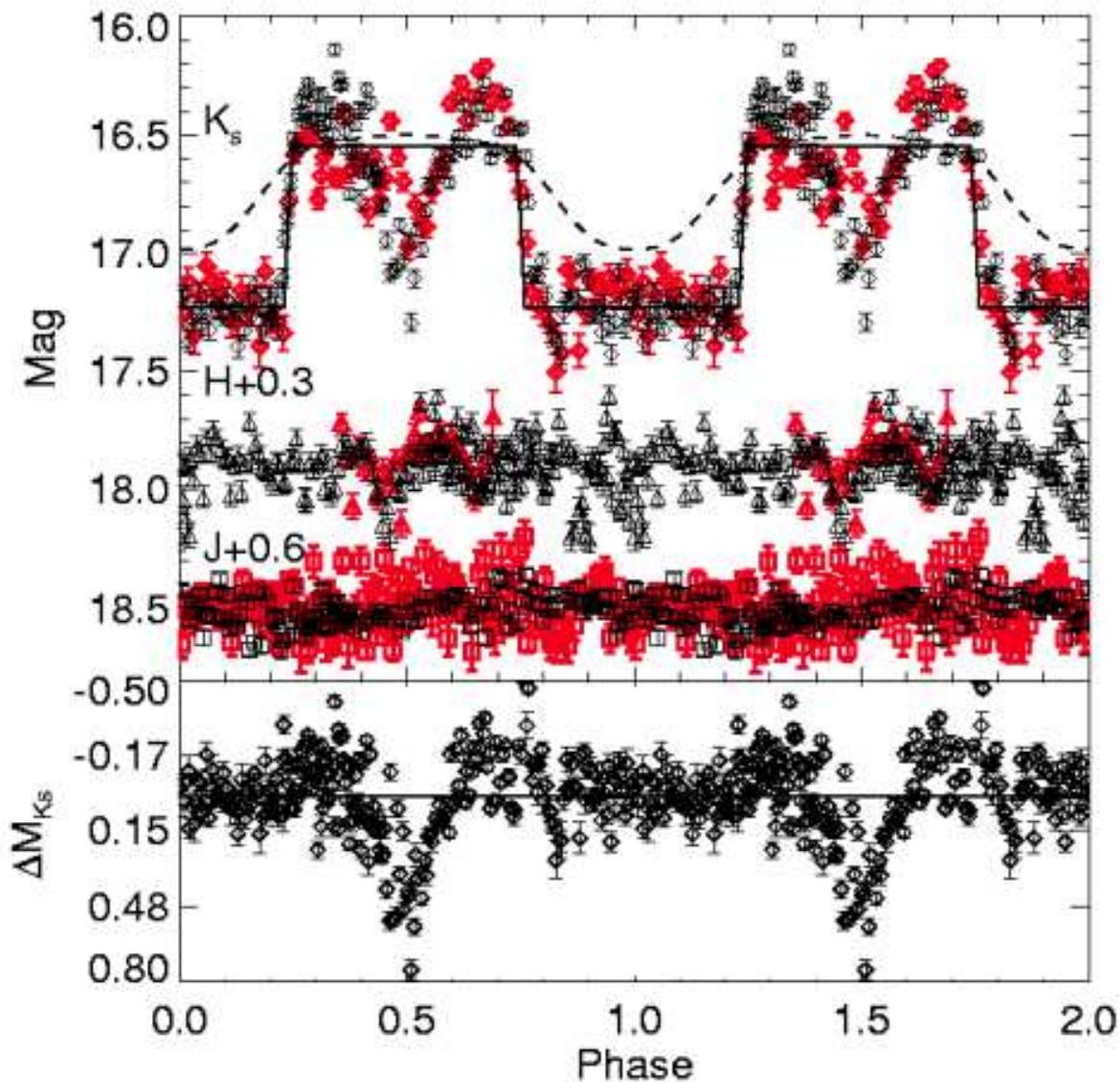}
\caption{\label{fig:lightcurve} Phased $J, H$, and $K_s$ lightcurves
of SDSS~1212 following Equation \ref{eq:eph}. Red points are from our
WIRC observations.
The solid line in the $K_s$ lightcurve
corresponds to a cyclotron spot model. The model
assumes a spot with a radius of 0.03R$_{WD}$ at the equator with a total
flux ratio to the white dwarf of 1.5.  The dashed line is an irradiated
companion model with $T_{hot}/T_{cool}$=1.4 and T$_{cool}$=1000~K.  The bottom plot shows the
residuals from the spot model.}
\end{figure}

We attempted to model the $J, H$, and $K_s-$band light curves of SDSS 1212
with WD2003 adopting various irradiation scenarios with effective
temperatures of the secondary down to 600K and assuming that the 
secondary co-rotates with the white dwarf in its orbit. 
The values of the orbital
period and initial epoch of the light curves were adopted from equation
1. We assumed a mass of $0.6\; \rm M_{\sun}$, a radius of $0.012\;\rm
R_{\sun}$ and a $T_{eff}$ of 10,000K for the white dwarf. The mass of
the companion was set to $0.06\; \rm M_{\sun}$, and the radius to the
size of its Roche lobe, although there is no sign of undergoing
accretion, as reported by \citet{schmidt05}. The values used for the
albedos and the limb darkening and gravity brightening coefficients are
the same as those adopted by \citet{harrison03} in their irradiation
models of EF Eri. The model providing the closest resemblance to the
light curves corresponds to a $T_{eff}$ = 1000K brown dwarf with a 1400K
irradiated side. That model is represented by the dashed line in Figure
\ref{fig:lightcurve}. Clearly, irradiation alone cannot reproduce the observed
light curves, since the irradiation curves produce a smooth rounded drop in 
flux which does not match the sharpness and the depth of the drop. 

In the second scenario, cyclotron emission associated with the
magnetic pole region of the white dwarf causes excess flux in
$K_s$.  Other magnetic white dwarf binaries, such as EF Eri, exhibit
strong cyclotron features even in the absence of accretion features
\citep{harrison04}.  We postulate that the majority of the cyclotron
emission is confined to a spot near the magnetic pole of the white dwarf
and is obscured by the white dwarf when it rotates out of view.  Given a
total excess above the white dwarf and its companion of 0.08 mJy, the 
cyclotron emission would have an approximate luminosity of $\sim3\times10^{27}$~ergs s$^{-1}$.  This is a lower limit on the total luminosity
 caused by accretion of a wind from the companion and the inferred mass accretion rate would be 4$\times10^{-16}$M$_{\odot}$ yr$^{-1}$, assuming
a white dwarf mass of 0.6\msun, and a white dwarf radius of 0.01 R$_\odot$.  
Assuming a magnetic field strength of 13 MG, the cyclotron emission
observed at the wavelength of 2.06~\micron\ corresponds to a harmonic
number of 4.  Cyclotron emission should be present at
4.1 and 8.2 \micron\, accessible in {\em Spitzer's} IRAC channels 2 and 4.

We have generated a simple geometric model of a spot on the white
dwarf to model the observed light curve.  Our model includes the sum of
the $K_{s}-$band white dwarf+companion flux plus the flux from the spot
as a function of spin phase.  We assume that the spot is not affected by
limb darkening relative to the disk of the white dwarf. The radius of
the spot, the flux ratio of the spot to the white dwarf+companion, and
the latitude of the spot's position are free parameters. The results are
mostly insensitive to spot radius or latitude assuming that the spot is
close to the equator and the spot radius is $<$10\% of the white dwarf
radius.  We assume a flux ratio of spot to white dwarf of 1.5, which is
derived from the difference in flux due to the dip compared to that
expected from the white dwarf alone.  The solid line in Figure
\ref{fig:lightcurve} shows our results for a model with a spot of radius 0.03
R$_{WD}$, located at the equator. The bottom diagram shows the residuals
from this model. Some structure still remains visible in those
residuals, e.g. the dip around phase 0.5, which could
be caused by beaming of the cyclotron emission,
self-shielding of the cyclotron emission column
 that presents less surface area as it points
towards Earth, or an eclipse of the emission region by the companion.
 Overall, our simplistic cyclotron emission model
provides a good first order approximation to the $K_s$ light curve.

\section{Conclusions}
\label{sec:conclusions}

We report the detection of the brown dwarf companion to the magnetic
white dwarf SDSS 121209.31+013627.7, whose presence was inferred by the
detection of a variable $H_{\alpha}$ emission line by
\citet{schmidt05}.  We provide a more accurate period for the system of
0.061$\pm$0.001 days.  No eclipses are apparent, however our data constrain
the inclination of the system to $\lesssim 78.5$ deg.  We are only able
to place an upper limit of L7 to the spectral type of the companion,
given that the $K_s-$band photometry
appears to be contaminated by an additional source. We propose cyclotron
radiation as the most likely cause of the extra emission, although our
hypothesis requires phase-resolved infrared spectroscopy to be
confirmed. Spectroscopy will reveal the presence or absence of cyclotron
radiation peaks and determine the spectral type of the companion. It is
possible that high-precision $J$ and $H$-band photometry 
may reveal a shallow eclipse
or other subtle structure in the light curves.  SDSS~1212 shows no evidence of ongoing accretion from Roche-lobe
overflow, and may differ from similar objects, like EF Eri, by being a 
pre-polar system \citep{schmidt05}.

Finally, high resolution optical spectra of
SDSS~1212 will better measure the radial velocity variations of the
H$_\alpha$ line and link the variations with those observed in $K_s$.  
Measuring the radial velocity of the white dwarf through its
hydrogen lines will be difficult given
the high probability that changes in the projected mean field strength
will erase any signature from the companion.
High resolution spectroscopy will also 
provide a direct measure of any material that
is accreted onto the white dwarf evidenced by the presence of weak
metal lines.  Magnetic white dwarf binaries can efficiently capture
the stellar wind of their companions and radiate the accretion energy
entirely in cyclotron emission \citep{schmidt05b,webbink05}.
Metal lines in DAZ white dwarfs have proved useful for
measuring low levels of mass accretion onto isolated white dwarfs and
white dwarfs with close companions and would provide an independent 
measure of the material accreted by the white dwarf.
\citep{holberg97,koester97,zuckerman03,debes06}.

\acknowledgments{}

We thank our referee Gary Schmidt for constructive suggestions for this
manuscript.  We thank Larry Petro for insightful discussion regarding
cyclotron spots and accretion columns for magnetic white dwarfs.
We also thank
Sara Seager and Hannah
Jang-Condell for their helpful suggestions. 
M.~L-M. acknowledges research and travel support from the Carnegie
Institution of Washington through a Carnegie
Fellowship. A.~Z.~B. acknowledges research and travel support from the
Carnegie Institution of Washington through a Vera Rubin Fellowship. 
 Funding for the SDSS and SDSS-II
has been provided by the Alfred P. Sloan Foundation, the Participating
Institutions, the National Science Foundation, the U.S. Department of
Energy, the National Aeronautics and Space Administration, the Japanese
Monbukagakusho, the Max Planck Society, and the Higher Education Funding
Council for England. The SDSS Web Site is http://www.sdss.org/



\clearpage

\begin{deluxetable}{cccc}
\tabletypesize{\footnotesize}
\tablecolumns{4}
\tablewidth{0pc}
\tablecaption{\label{tab:photometry} Calibration Stars}
\tablehead{ \colhead{Name} & \colhead{J} & \colhead{H} & \colhead{K$_s$} \\
 \colhead{} & \colhead{(mag)} & \colhead{(mag)} & \colhead{(mag)}}
\startdata
2MASS 12120989$+$0135259 & 16.23$\pm$0.12 & 15.89$\pm$0.17 & 15.54$\pm0.23$\\
2MASS 12120582$+$0135157 & 15.61$\pm$0.08 & 14.99$\pm$0.07 & 14.95$\pm0.14$\\
2MASS 12120408$+$0135365 & 15.65$\pm$0.07 & 15.05$\pm$0.06 & 14.78$\pm0.12$\\
2MASS 12121658$+$0137042 & 14.95$\pm$0.04 & 14.29$\pm$0.04 & 14.14$\pm0.07$\\
2MASS 12121667$+$0135257 & 14.45$\pm$0.03 & 13.86$\pm$0.02 & 13.80$\pm0.06$\\
SDSS J121205.54$+$013720.2 &-- &-- &-- \\
\enddata 
\end{deluxetable}

\clearpage

\begin{deluxetable}{cccc}
\tabletypesize{\footnotesize}
\tablecolumns{4}
\tablewidth{0pc}
\tablecaption{\label{tab:stub} $JHK_s$ Photometry}
\tablehead{ \colhead{HJD} & \colhead{Magnitude} & \colhead{Uncertainty} & \colhead{Band}}
\startdata
2453781.82495  &  17.889  &  0.041  &   J \\
2453781.82568  &  17.956  &  0.041  &   J \\
2453781.82642  &  17.877  &  0.041  &   J \\
\enddata 
\tablecomments{Table 2 is available in its entirety in the electronic
version of the Journal. A portion is shown here for guidance regarding
its form and content.}
\end{deluxetable}

\clearpage

\clearpage

\end{document}